**Voltage-controlled magnetic anisotropy under the electronic structure modulation in quantum wells**

Qingyi Xiang[1,2], Yoshio Miura[1], Muftah Al-Mahdawi[1], Thomas Scheike[1], Xiandong Xu[1], Yuya Sakuraba[1], Shinya Kasai[1], Zhenchao Wen[1], Hiroaki Sukegawa[1], Seiji Mitani[1,2], and Kazuhiro Hono[1,2]

1. *Research Center for Magnetic and Spintronic Materials, National Institute for Materials Science (NIMS), Tsukuba 305-0047, Japan*
2. *Graduate School of Pure and Applied Sciences, University of Tsukuba, Tsukuba 305-8577, Japan*

**Abstract**

Voltage-controlled magnetic anisotropy (VCMA) offers an emerging approach to realize energy-efficient magnetization switching in spintronic devices such as magnetic random access memories (MRAMs). Here, we show that manipulating the condensed states, *i.e.*, introducing quantum well (QW) can significantly influence the VCMA in a Cr/Fe-QW/MgAl$_2$O$_4$ based magnetic tunnel junction (MTJ). Only for the MTJ with an even number of Fe atomic layers, we observed a novel A-shaped VCMA curve for a particular QW state, where magnetic anisotropy energy (MAE) reaches a local maximum at zero bias and reduces when applying both positive and negative bias, *i.e.*, a novel bi-polar VCMA effect. Our *ab initio* calculations demonstrate that the QW states give an additional contribution to perpendicular magnetic anisotropy (PMA), which can explain not only the A-shaped VCMA but also the Fe-layer-number parity dependence of VCMA. The present study suggests that the QW-modulated VCMA should open a new pathway to design VCMA-assisted MRAM.

*Introduction*

Manipulation of magnetism through electrical current and voltage plays a crucial role to realize practical spintronic devices with high speed, low power-consumption and non-volatile properties, such as magnetic random access memory (MRAM).[1–6] The perpendicular magnetic anisotropy (PMA), arising from the spin-orbit interaction (SOI),[7–9] is of the extreme importance, which ensured the scalability and thermal stability of magnetic tunnel junction (MTJ) cells for MRAM.[10] However, a larger PMA energy density always leads to a more required switch energy, either spin-transfer-torque (STT) [11–14] or spin-orbit-torque (SOT)[4,15,16], resulting in sizeable switching current, along with energy-wasting and Joule heating issues, as the nature of PMA is the energy barrier to prevent magnetization to switch. Thus, developing a novel method that can manipulate the magnetization without large electric current such as the voltage-controlled magnetic anisotropy (VCMA) is strongly desired.[3,17–20]

VCMA in an all-solid film stack is firstly observed in Fe-MgO based MTJs.[3] Fast manipulation of magnetism down to 0.1 ns was realized,[21] along with the bistable magnetization switching [18,22] and ferromagnetic resonance excitation.[19,23] Considering that the SOI is regarded as the origin to explain the magnetic anisotropy,[7,8] it is believed that the electric field generated at the ferromagnetic layer/oxide interface plays a significant role in modulating the electronic structure, such as charge accumulation/depletion,[17,24,25] an electric-field-induced magnetic dipole[26] or the Rashba effect.[27]

Despite the mechanism, a large VCMA coefficient is required to realize the practical magnetization switching. In an Fe/MgO bilayer structure, a VCMA coefficient up to ~300 fJ/Vm was observed.[28–30] However, the coefficient is expected to be over 1000 fJ/Vm for practical memory applications. Meanwhile, all the VCMA behaviors reported so far are monodirectional, which means the anisotropy energy is only reduced when applied bias in a particular direction, while undesired enhancement occurs when opposite bias is applied. In order to use the VCMA effect to practical voltage-controlled devices, development of bidirectional VCMA is required.

Here, we show a new concept to modulate VCMA with bias-bidirectional change at the interface by modulating the SOI through introducing QWs in an MTJ pillar. In a QW-MTJ pillar, the resonant states in a ferromagnetic metal layer such as Fe are generated due to the quantum confinement of the density of states (DOS). Such DOS modulation is believed to lead to a SOI modification, which can affect the magnetic anisotropy behavior.[31] It was reported that magnetic anisotropy energy (MAE) shows oscillatory changes depending on the Fe film thickness due to the minority-band quantum well formation in HM(heavy metal)/Fe/HM structures.[32,33] However, it is difficult to employ these non-MTJ structures in practical spintronics applications due to the absence of large PMA, detectable magnetoresistance (MR), and insulated interfaces to apply external electric fields. Hence, in this work, we demonstrated a conventional MTJ structure, where a QW is introduced by precisely interface engineering, to investigate the QW-VCMA effect with.

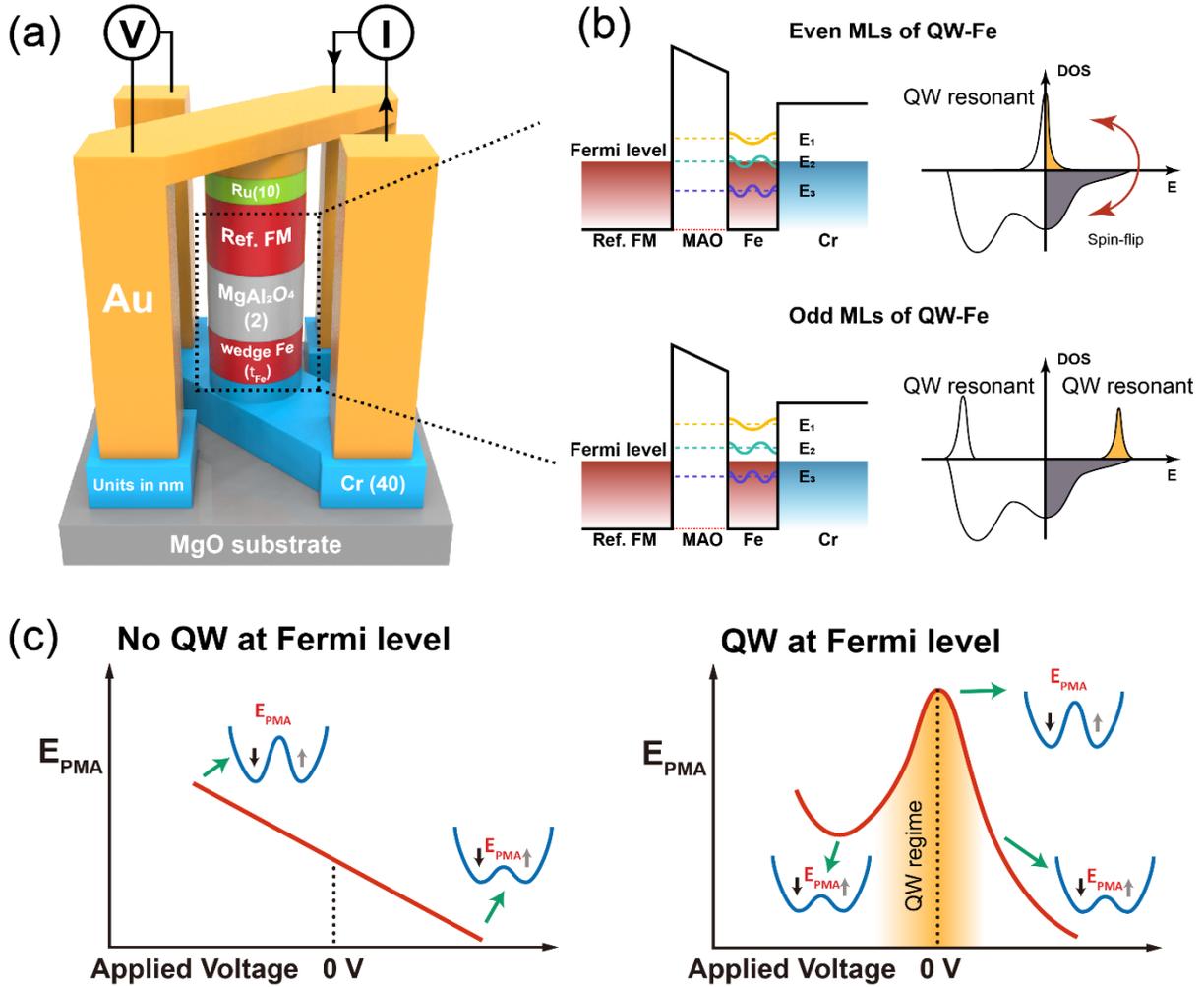

**Figure 1 Concepts of QW modulation on VCMA:** (a) QW-MTJ device structure. "Wedge-Fe" provides the QWs and perpendicular magnetic anisotropy. (b) Energy diagrams and corresponding DOS with modulation introduced by the QW effect. For even Fe-MLs-case with QW-Fe where resonant state located near Fermi level, and the spin-flip term can play a dominant role in PMA energy. Moreover, for the odd-MLs case where resonant state located far away from the Fermi level, there is no spin-flip term contributed to PMA energy. (c) PMA energy $E_{PMA}$ as a function of bias voltage for the (left) odd-MLs case and (right) even-MLs case. PMA energy and corresponding energy barrier to magnetization switch can be designable by applying bias voltage via formed resonant states. Especially, an "A" shaped VCMA curve is expected when one QW state is located near zero bias.

*Experiments*

We prepared MgAl$_2$O$_4$-based MTJ devices using a Cr/ultrathin Fe/oxide epitaxial structure to create stable QWs,[34–38] as shown in Figure 1 (a). The MTJ stacks consist of MgO substrate/Cr (40 nm)/ultrathin-Fe (nominal $t_{Fe}$, varies from 4 to 7 monolayers (MLs))/MgAl$_2$O$_4$ (2

nm)/reference ferromagnetic layer (ref. FM)/Ru cap. The top ref. FM is a thick and in-plane magnetized CoFeB or Fe layer. The reason for the selection of the MgAl$_2$O$_4$ barrier to realize fully-lattice-matched MTJ structure.[38,39]

The core structure of Cr/Fe/MgAl$_2$O$_4$ with atomic layer-by-layer growth provides very strong perpendicular magnetization due to Fe-O orbital hybridization,[8,9,39] and a "quasi" QW layer due to the quantum confinement.[35,36,38] As shown in Figure 1 (b), the band mismatch among Cr, Fe, and MgAl$_2$O$_4$ of $\Delta_1$ Bloch states create a QW in the Fe majority spin band (i.e., Fe-$\Delta_1\uparrow$). The formed QWs vary their locations depending on the Fe thickness. Strong modulation in the transport happens when the QW position matches with the Fermi level ($E_F$) of ref. FM by a bias voltage applied to MgAl$_2$O$_4$. For very thin (< 10 MLs) QW-Fe case, the even-MLs-QW has a resonant state near $E_F$ while an odd-MLs-QW does not have due to the inherent band dispersion of Fe-$\Delta_1\uparrow$.[35,36,38] Such a near-$E_F$ resonant state can significantly influence the magnetic anisotropy by the hybridization between majority and minority bands near the Fe/Oxide interface.[8] This designable QW via thickness control makes it possible to obtain a large PMA energy density at zero-bias, which can be reduced by bias voltage application; *i.e.*, reduction of the energy barrier by the QW-modulated-VCMA that is desirable for switching process as schematically shown in Figure 1 (c).

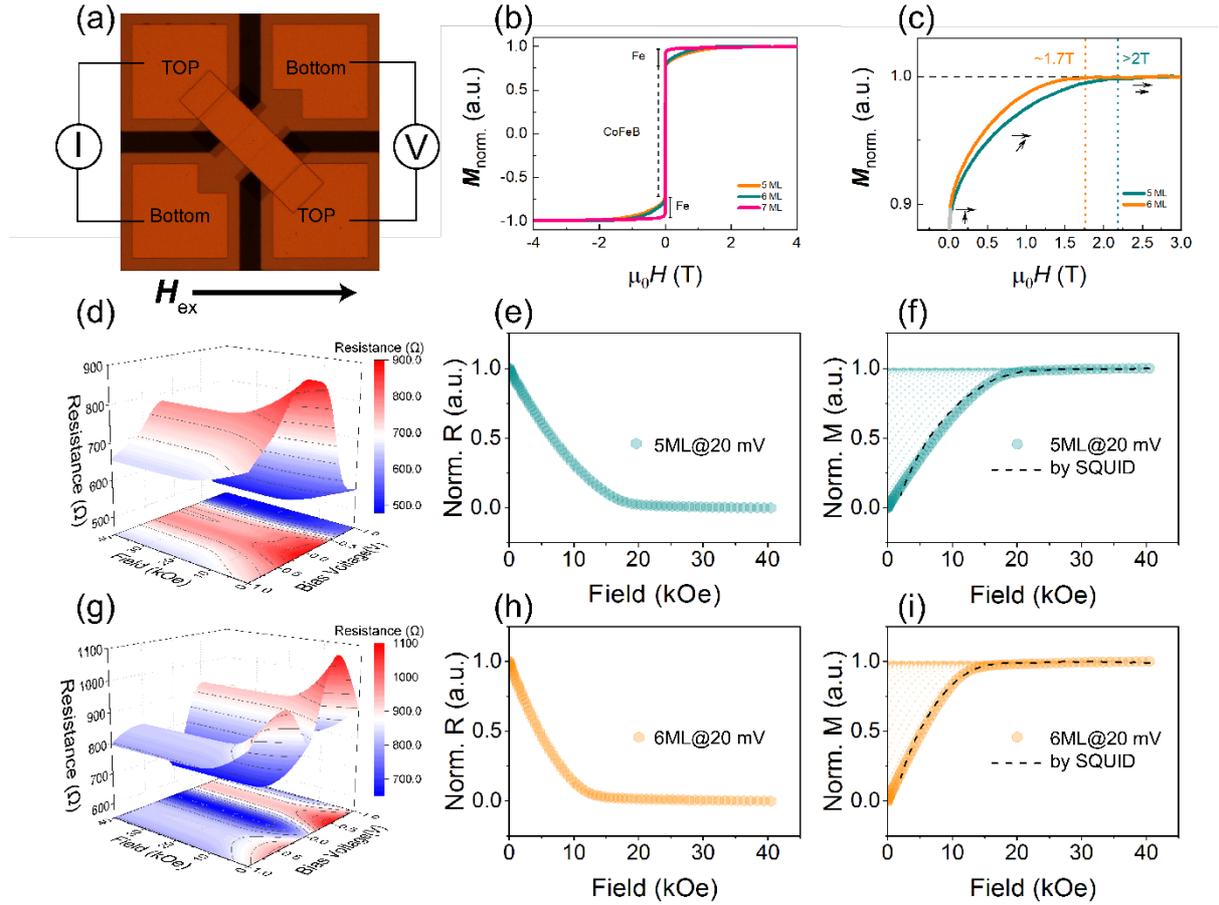

**Figure 2 Measurements for patterned MTJs** (a) MTJs measured by 4-probe measurement in this study, the external field is applied in-plane direction(hard axis of the QW-Fe layer). (b) the ***M-H*** curves measured before patterning for 5-, 6-, and 7-ML Fe MTJ films, respectively. (c) the subtracted ***M-H*** curves contributed from the QW-Fe layer, which indicted large anisotropy filed for 5- and 6-ML samples. Collected resistance-voltage-field maps show resonant features for (d) 5-ML and (g) 6-ML Fe MTJs. (e) and (h) are sliced resistance-field curves for 5- and 6-ML MTJs under 20 mV bias, which can be converted into ***M-H*** curves as (f) and (i) show, agreeing well with SQUID measured results as dashed lines show.

To verify the QW effect on VCMA, MTJs were prepared and measured by both magnetic and electrical methods to evaluate their PMA properties (See *Methods*). Figure 2 (a) shows the MTJ device used in the electrical measurements. Magnetoresistance was measured from the device as a function of the applied field for various applied bias voltages with high resolution up to 5 mV, which can be converted into an ***M-H*** curve (see *Methods*). To confirm the electrical measurement result, a conventional magnetic property measurement is also performed using SQUID before microfabrication. The ***M-H*** curves for ultrathin QW-Fe measured in-plane direction are shown in Figure 2 (b), where the anisotropy field ($H_k$) for 5- and 6-ML films are

~2T and ~1.7T, respectively. The dashed lines in Figure 2 (d) and (e) are the SQUID results, which agree well with the electrical measurement result with a small applied voltage. The PMA energy densities were calculated and analyzed for the bias dependence of PMA energy densities. Meanwhile, the *dI/dV* spectrum for each MTJ is also collected, since the differential conductance spectrum contains the DOS information of the ferromagnetic layer.[34]

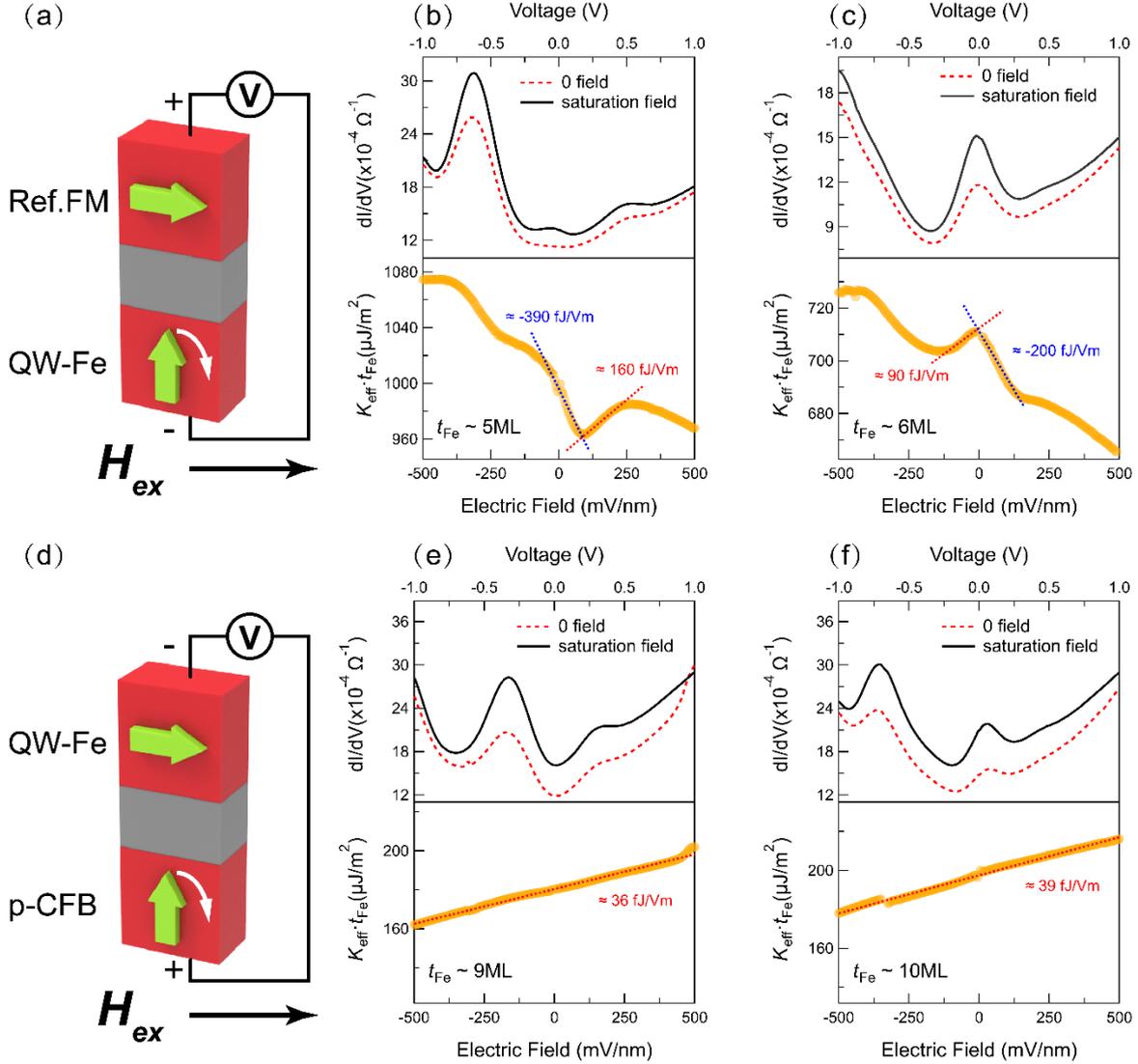

**Figure 3 QW effect on VCMA behaviors.** The *dI/dV* spectrum and VCMA curves for VCMA measured for (a) QW-Fe with (b) 5-ML and (c) 6-ML Fe MTJs. The peak of VCMA curve near 0 bias in (c) indicating the QWs near Fermi level modulated the PMA energy density. As reference, the *dI/dV* spectrum and VCMA curves are also measured for (e) the perpendicular CoFeB with in-plane Fe, where QW resonant observed in the in-plane Fe with (e) 9-ML and (f) 10-ML.

Figure 3 shows the voltage dependence of both differential conductance, i.e., *dI/dV*, and effective area anisotropy energy densities, *i.e.*, $K_{\text{eff}} \cdot t_{\text{Fe}}$, in the MTJs with QW-Fe. The QW-Fe layer with PMA are examined as Figure 3 (a) shows, where the Fe thickness are 5-ML and 6-ML, respectively. In the upper field of Figure 3 (b) and (c), the peaks of *dI/dV* indicate the existence of QWs in the structures[35,36,38]. For 5-ML Fe MTJ, the two resonant peaks locate at around -0.585 V and +0.510 V, respectively. Meanwhile, the resonant position occurs near zero bias for 6-ML MTJs, as shown in Figure 3 (c). Owing to the modified DOS of the QW-Fe layers in the MTJs, the VCMA curves show a unique feature rather than the typical linear behavior in poly-CoFeB/MgO[40]. For the 5-ML MTJ, a local minimum is observed at near 100 mV/nm, which is similar to previous reports.[28,29]

Furthermore, in this work, the high-resolution measurement presents a continuous observation of how anisotropy energy varies depending on applied bias voltage. The observation clearly indicates that the VCMA features of ultrathin-Fe/oxide originate from the combined effect of the electric field and the modification on DOS of ultrathin-Fe. Here, the former terms present typically a liner behavior of VCMA[21,40], which is believed to be contributed from the electron accumulation and depletion. The latter term is the new finding in this study, which is contributed from the QWs created in the ultrathin-Fe layer, as significant fine structures can be observed in Figure 3 (b) and (c). For the 5-ML case, the resonant states located at a high bias range, where the DOS modification is rather complicated and un-predictable. The VCMA coefficient can reach up to around -390 fJ/Vm at zero bias and quickly changes its sign and reach to around 160 fJ/Vm.

For further proving the origin of this anomalous VCMA behavior originated from QW resonant states of the Fe, another group of MTJs, where PMA arised from perpendicular CoFeB, are measured with existence of QW states provided by in-plane Fe as shown in Figure 3 (d), with thickness of 9- and 10-ML, respectively. Note that the polarity definition is reserved with MTJs in Figure 3 (a), for uniformity of *dI/dV* spectrum polarities. Typical linear VCMA behaviors are observed no matter where QWs located as Figure 3 (e) and (f) shows. It strongly suggested the VCMA behavior is definitely modulated by the QW states inside the Fe.

Figure 3 (c) shows how QWs affect MAE for the 6-ML case. The PMA energy reaches its local maximum at near-zero bias where the QWs located, indicating that PMA is enhanced by the DOS modulation. Besides the peak at near-zero bias, the VCMA coefficient is around 90 fJ/Vm, and -200 fJ/Vm with a negative and a positive bias, respectively. Thus, when bias voltage is applied, the PMA energy reduces for regardless of the polarity, *i.e.*, the VCMA is bi-polar. Note that PMA energy reduces one direction and increases in the opposite direction in the typical VCMA system.[28,30,40] When considering the VCMA combined with the widely used STT switching method, finding the bi-polar VCMA is a significant advantage. Combining this bipolar VCMA behavior, the switching current is significantly reduced for both Parallel(P) to Antiparallel(AP) and AP to P switching, which is impossible for a typical VCMA system since there is always one direction that the PMA energy increases with applied bias.

*Discussion*

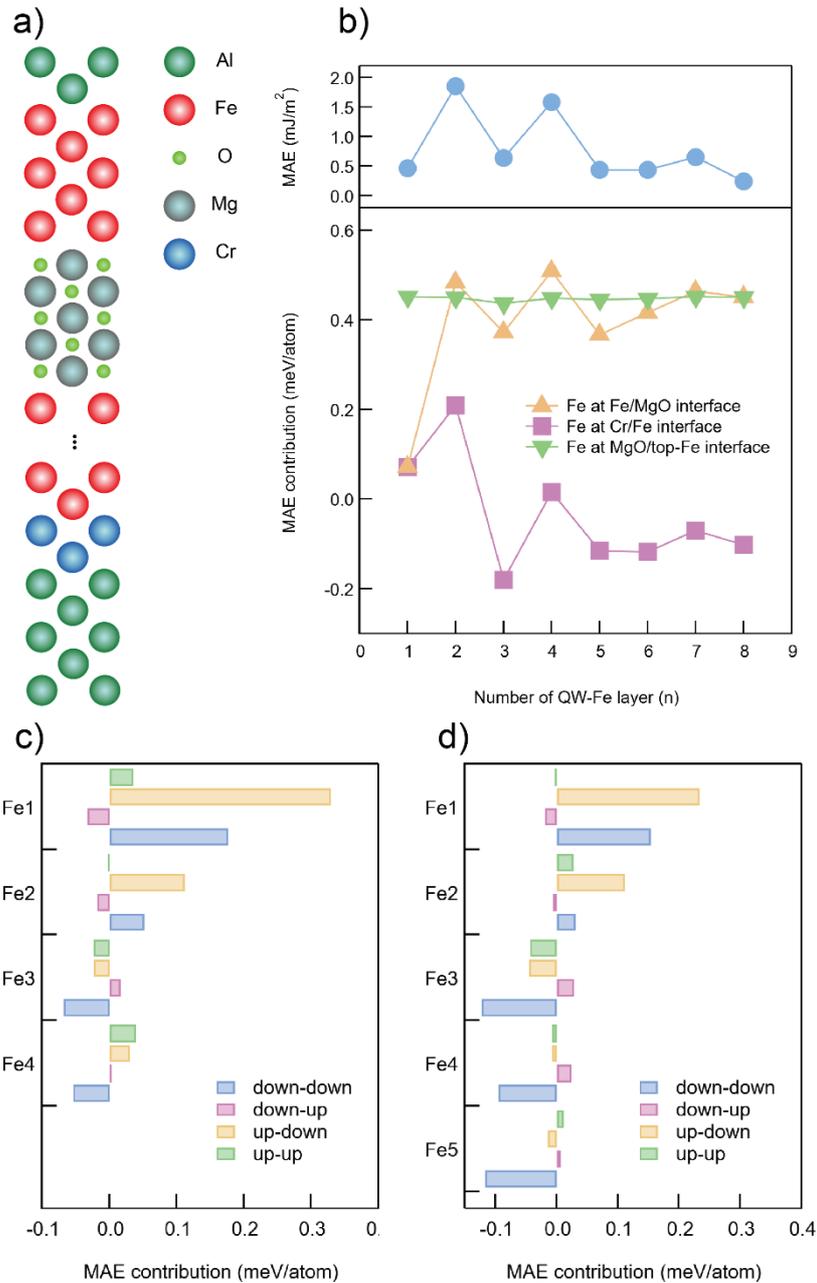

**Figure 4** *ab initio* calculation analysis for spin-resolved contributions to PMA energies. (a) The simulated model used for *ab initio* calculation in this study; (b) calculated MAE of the whole structure(upper) and resolved-channels of contribution from Fe/MgO, Cr/Fe and MgO/top-Fe interfaces(lower). The major contributor to the oscillation is the Fe/MgO interface. (c) and (d) shows the layer-resolved MAE contribution for even(4-ML) and odd(5-ML) layers of the QW-Fe, where the major contribution is from the spin-flip term, *i.e.,* up-down term.

In order to clarify the mechanism of the QW effect on VCMA and PMA, *ab initio* calculations (see *Methods*) were performed using a simulated structure, as shown in Figure 4 (a). In the upper part of Figure 4 (b), we show the MAE as a function of the number of the Fe layer (*n*), *i.e.,* QW width. It is found that MAE of the super-cell shows oscillating behavior with respect to the number of Fe layer, associated with the QW resonant position oscillations[38]. The MTJs with even number of Fe layer tend to show larger PMA, while these with odd number of Fe layers tend to show smaller PMA. It is consistent with the experimental results, where PMA is enhanced when QW forms near the Fermi level (zero bias voltage). Moreover, the detailed atomic contribution to the PMA energy is analyzed in Figure 4 (b) for the interfacial Fe atoms at Cr/Fe interface, at the bottom Fe/MgO(001) interface and at the upper Fe/MgO(001) interface, respectively, as a function of the number of Fe layers. It is clear that the interfacial Fe atom at the bottom Fe/MgO(001) interface shows a main contribution to the oscillation. The interfacial Fe atom at the Cr/Fe interface also shows the oscillation; however, the contribution to PMA is relatively small. Furthermore, the interfacial Fe atom at the upper Fe/MgO(001) does not show any oscillation of the PMA energy against the Fe thickness, which suggests the QWs between Cr/Fe/MgO(001) play a significant role in the oscillation of the PMA energy.

For further understanding, we selected two samples (with and without QWs near Fermi level) to analyze how the QW affects the MAE in detail. The spin-resolved MAE $\Delta E$(up-up), $\Delta E$(down-down), $\Delta E$(up-down) and $\Delta E$(down-up) in the second order perturbation of SOI are calculated from Equation (3) at each atomic site of interfacial Fe atom for even (*n*=4) and odd (*n*=5) Fe layers respectively. It is found that the spin-flip $\Delta E$(up-down) of the interfacial Fe atom (Fe1) mainly contributes to the PMA at the bottom Fe/MgO(001) interface. This can be attributed to the formation of the QWs of the majority-spin $\Delta_1$ wavefunction such as $d_{3z^2-r^2}$ in the Cr/Fe/MgO. The $d_{3z^2-r^2}$ state in the majority-spin state provides the spin-flip matrix elements of $L_x$ with the $d_{yz}$ state in the minority-spin state, leading to the enhancement of the PMA. Furthermore, the contribution of the spin-flip term $\Delta E$(up-down) at the bottom Fe/MgO interface is strongly enhanced in the case of the even Fe layers as compared with the odd Fe layer. This means that the spin-flip term $\Delta E$(up-down) of the interfacial Fe atoms is the origin of the oscillation of the PMA in the Cr/Fe(n)/MgO/Fe. The peak of projected LDOS of interfacial Fe $d_{3z^2-r^2}$ orbital at Cr/Fe and Fe/MgO interfaces provides the QWs due to interfacial bonding between Fe $d_{3z^2-r^2}$ and O $p_z$. Furthermore, the QWs of the majority-spin $\Delta_1$ is close to the Fermi level for MTJs with the even Fe layers, which yields the significant matrix elements of $L_x$ between the majority-spin $d_{3z^2-r^2}$ and the minority-spin $d_{yz}$. These calculation results explain how the QW effect brings an impact on the VCMA behavior through the spin-flip term in the second-order perturbation of the SOI.

*Conclusions*

In summary, we prepared perpendicular MTJs with QWs in atomically controlled ultrathin Fe layer. By introducing the QWs, the DOS of the ferromagnetic layer is modified leading to a novel bi-polar VCMA behavior, *i.e.*, A-shaped VCMA curve caused by QW resonant at near-zero bias voltage. The VCMA in odd and even number of Fe layers gives a different response to the bias voltage due to the shifted resonant bias associated with the QW width. Our *ab initio* calculations suggest that the QWs contribute to the magnetic anisotropy energy via the spin-flip term in the second-order perturbation of the SOI. Such DOS engineering by introducing QW structure to manipulate VCMA provides a new thought for electrical control of magnetism. Especially the A- shaped VCMA curve observation, *i.e.* the bi-polarity of VCMA, broads the possibilities to utilize the VCMA effect to magnetization switching in voltage controlled MRAM.

*Methods*

**Sample preparation**

A fully epitaxial MgO (5 nm)/Cr (30 nm)/Fe-QW ($t_{Fe}$: 0.45−1.25 nm)/MgAl$_2$O$_4$ (2 nm)/Fe (10 nm)/Ru (15 nm) was prepared on a single crystalline MgO(001) substrate. All the layers, except Ru capping layer, were deposited by electron beam evaporation under base pressure of less than $1\times10^{-8}$ Pa. The substrate was firstly annealed at 800°C for degassing and cleaning surface. Then the 5 nm MgO was deposited on the substrate at 450°C as a seeding layer. The Cr, QW-Fe, and MgAl$_2$O$_4$ layers were deposited at 150°C where the post annealing was performed at 800°C, 250°C, and 400°C respectively. The wedge-shaped Fe layer was deposited with a step of around 0.02 nm using a linear motion shutter equipped between the Fe source and substrate. The top reference ferromagnetic layer was deposited at RT without post-annealing to prevent the formation of perpendicular anisotropy at the MgAl$_2$O$_4$/FM interface. Finally, the stacks were capped with Ru layer via RF magnetron sputtering with the process pressure of 0.4 Pa at RT without post-annealing. The prepared film was patterned into ellipse junctions of 5×10 μm$^2$ in size by a conventional micro-fabrication method using Ar ion etching and photolithography.

**VCMA measurements**

Two independent methods examined the PMA properties. Before microfabrication of MTJs, the thin films were measured by superconducting quantum interference device(SQUID). Note that the films have orthogonal easy axis provide by Ref. FM and QW-Fe. The contributions from QW-Fe is subtracted from out-of-plane direction to evaluate the $H_k$.

Then MTJs were measured for magnetoresistance (MR) and the current-voltage (*I-V*) curves using a dc 4-probe method, with bias voltage up to ±1 V. The differential conductance (*dI/dV*) spectra were mathematically calculated from the *I-V* curves. The magnetic field up to 2 T was applied in the plane of the film for the measurement at RT. Since the ref. FM layer has an in-plane easy axis, the magnetization of the ref. layer is saturated along the external field immediately, and the QW-Fe layer with out-of-plane magnetization will gradually rotate its magnetization with field increasing. As Figure 3(c) and (e) shows, owing to the orthogonal configuration of the two magnetizations, the magnetoresistance, $R(\theta)$, will gradually decrease when applied field increasing as it is given by[28]:

$$R(\theta) = \frac{R_{90} R_p}{R_p + (R_{90} - R_p) \cdot \cos\theta}, (1)$$

A standard method to evaluate the PMA is established by monitoring the resistance variation with the applied field, and convert the resistance change into moment change along the in-plane direction as:[28]

$$\frac{M_{in-plane}}{M_s} = \cos\theta = \frac{R_{90} - R(\theta)}{R(\theta)} \frac{R_p}{R_{90} - R_p}, (2)$$

**First-principles calculation**

A first-principles calculations for Al/Cr/Fe/MgO/Fe(001) super-cell is performed using the Vienna ab initio simulation package (VASP) [41]. We adopt the spin-polarized generalized gradient approximation (GGA) proposed by Perdew, Burke, and Ernzerhof for the exchange and correlation energy[42], in which wavefunctions are expanded in a plane-wave basis set, and behaviors of core electrons are described by the projector augmented wave (PAW) potential[43,44]. The simulated structure was Al (5)/Cr (2)/Fe (1-8)/MgO (5)/Fe (5)/Al (2) (thickness in MLs), where the number of Fe MLs (n) was varied from 1 to 8 MLs [Fig. 4(a)]. In the previous work[45], we confirmed that the origin of PMA of Fe/MgAl$_2$O$_4$(001) interface is consistent with that of Fe/MgO(001) interface. Thus, we consider MgO instead of MgAl$_2$O$_4$ in the barrier layer of the present study because of the reduction of the computational cost. The inter-layer distances of Al/Cr, Cr/Fe and Fe/MgO interfaces along the film normal are fully optimized. The Al layers were included as a non-polarized bath of $\Delta_1$ electrons, and the insertion of the Cr layer is used to control the formation of the QW. Cr 2ML is enough to conform QWRS for the $\Delta_1$ state of Fe in Cr/Fe/MgO(001). The PMA energy is determined by the force theorem[46], where the differences in the sum of energy eigenvalues between the magnetization oriented along the in-plane [100] and out-of-plane [001] directions are calculated with the SOI after reading the charge densities of the system calculated without SOI. The PMA energy is defined as positive for perpendicular magnetization, i.e., $E_{PMA} = E_{[100]} - E_{[001]}$. The k-point integration is performed using a modified tetrahedron method with Blöchl corrections[47] with 24×24×2 k-points in the first Brillouin zone of the super-cell. Furthermore, to discuss the relation between the PMA and the electronic density of states, we calculate the PMA energy $E^{(2)}_{PMA}$ of each site using the second-order perturbation of the SOI on the basis of the projection of wavefunctions of the VASP-PAW calculations onto the local atomic orbitals:[48,49]

$$E^{(2)}_{PMA} = -\sum_{k}\sum_{n'\sigma'}^{unocc}\sum_{n\sigma}^{occ}\frac{|\langle kn'\sigma'|H_{SO}|\rangle kn\sigma|^2}{\epsilon^{(0)}_{kn'\sigma'} - \epsilon^{(0)}_{kn\sigma}}, (3)$$

The details of the formulation of $E^{(2)}_{PMA}$ is described in ref.[48,49]. In this work, the SOI constants of Al, Cr, Fe, O and Mg are set as 20meV, 45meV, 50meV, 15meV and 10meV, respectively. The validity of these spin-orbit constants is confirmed by the qualitative agreement between the sum of $E^{(2)}_{PMA}$ for each site and the total PMA energy from the first-principles calculations.

**References**

1. Chappert, C., Fert, A. & Van Dau, F. N. The emergence of spin electronics in data storage. *Nat. Mater.* **6**, 813–823 (2007).

2. Ikeda, S. *et al.* Magnetic Tunnel Junctions for Spintronic Memories and Beyond. *IEEE Trans. Electron Devices* **54**, 991–1002 (2007).

3. Maruyama, T. *et al.* Large voltage-induced magnetic anisotropy change in a few atomic layers of iron. *Nat. Nanotechnol.* **4**, 158–161 (2009).

4. Liu, L. *et al.* Spin-Torque Switching with the Giant Spin Hall Effect of Tantalum. *Science* **336**, 555–558 (2012).

5. Kent, A. D. & Worledge, D. C. A new spin on magnetic memories. *Nat. Nanotechnol.* **10**, 187–191 (2015).

6. Waldrop, M. M. The chips are down for Moore's law. *Nat. News* **530**, 144 (2016).

7. Bruno, P. Tight-binding approach to the orbital magnetic moment and magnetocrystalline anisotropy of transition-metal monolayers. *Phys. Rev. B* **39**, 865–868 (1989).

8. Yang, H. X. *et al.* First-principles investigation of the very large perpendicular magnetic anisotropy at Fe|MgO and Co|MgO interfaces. *Phys. Rev. B* **84**, 054401 (2011).

9. Koo, J. W. *et al.* Large perpendicular magnetic anisotropy at Fe/MgO interface. *Appl. Phys. Lett.* **103**, 192401 (2013).

10. Watanabe, K., Jinnai, B., Fukami, S., Sato, H. & Ohno, H. Shape anisotropy revisited in single-digit nanometer magnetic tunnel junctions. *Nat. Commun.* **9**, 1–6 (2018).

11. Slonczewski, J. C. Current-driven excitation of magnetic multilayers. *J. Magn. Magn. Mater.* **159**, L1–L7 (1996).

12. Berger, L. Emission of spin waves by a magnetic multilayer traversed by a current. *Phys. Rev. B* **54**, 9353–9358 (1996).

13. Albert, F. J., Katine, J. A., Buhrman, R. A. & Ralph, D. C. Spin-polarized current switching of a Co thin film nanomagnet. *Appl. Phys. Lett.* **77**, 3809–3811 (2000).

14. Mangin, S. *et al.* Current-induced magnetization reversal in nanopillars with perpendicular anisotropy. *Nat. Mater.* **5**, 210–215 (2006).

15. Miron, I. M. *et al.* Perpendicular switching of a single ferromagnetic layer induced by in-plane current injection. *Nature* **476**, 189–193 (2011).

16. Kim, J. *et al.* Layer thickness dependence of the current-induced effective field vector in Ta|CoFeB|MgO. *Nat. Mater.* **12**, 240–245 (2012).

17. Nakamura, K. *et al.* Giant Modification of the Magnetocrystalline Anisotropy in Transition-Metal Monolayers by an External Electric Field. *Phys. Rev. Lett.* **102**, 187201 (2009).

18. Shiota, Y. *et al.* Induction of coherent magnetization switching in a few atomic layers of FeCo using voltage pulses. *Nat. Mater.* **11**, 39–43 (2012).

19. Nozaki, T. *et al.* Electric-field-induced ferromagnetic resonance excitation in an ultrathin ferromagnetic metal layer. *Nat. Phys.* **8**, 492–497 (2012).

20. Wang, W.-G., Li, M., Hageman, S. & Chien, C. L. Electric-field-assisted switching in magnetic tunnel junctions. *Nat. Mater.* **11**, 64–68 (2012).

21. Nozaki, T., Shiota, Y., Shiraishi, M., Shinjo, T. & Suzuki, Y. Voltage-induced perpendicular magnetic anisotropy change in magnetic tunnel junctions. *Appl. Phys. Lett.* **96**, 022506 (2010).

22. Kanai, S. *et al.* Electric field-induced magnetization reversal in a perpendicular-anisotropy CoFeB-MgO magnetic tunnel junction. *Appl. Phys. Lett.* **101**, 122403 (2012).

23. Zhu, J. *et al.* Voltage-Induced Ferromagnetic Resonance in Magnetic Tunnel Junctions. *Phys. Rev. Lett.* **108**, 197203 (2012).

24. Duan, C.-G. *et al.* Surface Magnetoelectric Effect in Ferromagnetic Metal Films. *Phys. Rev. Lett.* **101**, 137201 (2008).

25. Tsujikawa, M. & Oda, T. Finite Electric Field Effects in the Large Perpendicular Magnetic Anisotropy Surface Pt/Fe/Pt(001): A First-Principles Study. *Phys. Rev. Lett.* **102**, 247203 (2009).

26. Miwa, S. *et al.* Voltage controlled interfacial magnetism through platinum orbits. *Nat. Commun.* **8**, 15848 (2017).

27. Barnes, S. E., Ieda, J. & Maekawa, S. Rashba Spin-Orbit Anisotropy and the Electric Field Control of Magnetism. *Sci. Rep.* **4**, (2014).

28. Nozaki, T. *et al.* Large Voltage-Induced Changes in the Perpendicular Magnetic Anisotropy of an MgO-Based Tunnel Junction with an Ultrathin Fe Layer. *Phys. Rev. Appl.* **5**, 044006 (2016).

29. Xiang, Q. *et al.* Nonlinear electric field effect on perpendicular magnetic anisotropy in Fe/MgO interfaces. *J. Phys. Appl. Phys.* **50**, 40LT04 (2017).

30. Nozaki, T. *et al.* Highly efficient voltage control of spin and enhanced interfacial perpendicular magnetic anisotropy in iridium-doped Fe/MgO magnetic tunnel junctions. *NPG Asia Mater.* **9**, e451 (2017).

31. Chang, C.-H., Dou, K.-P., Guo, G.-Y. & Kaun, C.-C. Quantum-well-induced engineering of magnetocrystalline anisotropy in ferromagnetic films. *NPG Asia Mater.* **9**, e424 (2017).

32. Dąbrowski, M. *et al.* Oscillations of the Orbital Magnetic Moment due to $d$-Band Quantum Well States. *Phys. Rev. Lett.* **113**, 067203 (2014).

33. Li, J., Przybylski, M., Yildiz, F., Ma, X. D. & Wu, Y. Z. Oscillatory Magnetic Anisotropy Originating from Quantum Well States in Fe Films. *Phys. Rev. Lett.* **102**, 207206 (2009).

34. Nagahama, T., Yuasa, S., Suzuki, Y. & Tamura, E. Quantum-well effect in magnetic tunnel junctions with ultrathin single-crystal Fe(100) electrodes. *Appl. Phys. Lett.* **79**, 4381–4383 (2001).

35. Lu, Z.-Y., Zhang, X.-G. & Pantelides, S. T. Spin-Dependent Resonant Tunneling through Quantum-Well States in Magnetic Metallic Thin Films. *Phys. Rev. Lett.* **94**, 207210 (2005).

36. Niizeki, T., Tezuka, N. & Inomata, K. Enhanced Tunnel Magnetoresistance due to Spin Dependent Quantum Well Resonance in Specific Symmetry States of an Ultrathin Ferromagnetic Electrode. *Phys. Rev. Lett.* **100**, 047207 (2008).

37. Sheng, P. *et al.* Detailed analysis of spin-dependent quantum interference effects in magnetic tunnel junctions with Fe quantum wells. *Appl. Phys. Lett.* **102**, 032406 (2013).

38. Xiang, Q. *et al.* Realizing Room-Temperature Resonant Tunnel Magnetoresistance in Cr/Fe/MgAl$_2$O$_4$ Quasi-Quantum Well Structures. *Adv. Sci.* 1901438 (2019) doi:10.1002/advs.201901438.

39. Xiang, Q., Mandal, R., Sukegawa, H., Takahashi, Y. K. & Mitani, S. Large perpendicular magnetic anisotropy in epitaxial Fe/MgAl$_2$O$_4$(001) heterostructures. *Appl. Phys. Express* **11**, 063008 (2018).

40. Li, X. *et al.* Enhancement of voltage-controlled magnetic anisotropy through precise control of Mg insertion thickness at CoFeB|MgO interface. *Appl. Phys. Lett.* **110**, 052401 (2017).

41. Kresse, G. & Hafner, J. Ab initio molecular dynamics for liquid metals. *Phys. Rev. B* **47**, 558–561 (1993).

42. Perdew, J. P., Burke, K. & Ernzerhof, M. Generalized Gradient Approximation Made Simple. *Phys. Rev. Lett.* **77**, 3865–3868 (1996).

43. Blöchl, P. E. Projector augmented-wave method. *Phys. Rev. B* **50**, 17953–17979 (1994).

44. Kresse, G. & Joubert, D. From ultrasoft pseudopotentials to the projector augmented-wave method. *Phys. Rev. B* **59**, 1758–1775 (1999).

45. Masuda, K. & Miura, Y. Bias voltage effects on tunneling magnetoresistance in Fe/MgAl$_2$O$_4$/Fe(001)$ junctions: Comparative study with Fe/MgO/Fe(001) junctions. *Phys. Rev. B* **96**, 054428 (2017).

46. Weinert, M., Watson, R. E. & Davenport, J. W. Total-energy differences and eigenvalue sums. *Phys. Rev. B* **32**, 2115–2119 (1985).

47. Blöchl, P. E., Jepsen, O. & Andersen, O. K. Improved tetrahedron method for Brillouin-zone integrations. *Phys. Rev. B* **49**, 16223–16233 (1994).

48. Miura, Y., Tsujikawa, M. & Shirai, M. A first-principles study on magnetocrystalline anisotropy at interfaces of Fe with non-magnetic metals. *J. Appl. Phys.* **113**, 233908 (2013).

49. Miura, Y. *et al.* The origin of perpendicular magneto-crystalline anisotropy in L10–FeNi under tetragonal distortion. *J. Phys. Condens. Matter* **25**, 106005 (2013).

*Acknowledgments*

This study was partially supported by the ImPACT program of the Council for Science, Technology and Innovation (Cabinet Office, Government of Japan) and JSPS KAKENHI Grant Number 16H06332. Q.X. acknowledges the National Institute for Materials Science for providing a NIMS Junior Research Assistantship. We also thank T. Ohkubo for supporting TEM observation; M. Belmoubarik, T. Oda. M. Shirai, T. Nozaki and S. Yuasa for fruitful discussions.

**Author contributions**

Q.X. and S.M designed the experiments, Q.X. performed the sample preparation, measurement, and data analysis. Z.W., H.S., M.A., and T.S. contributed to sample preparation and microfabrication, Y.S and S.K. contributed to measurement set-up, Y.M. developed the theory, X.X. performed TEM observation, Q.X. and M.S. drafted the manuscript with discussion with all other co-authors. All authors commented on the manuscript.

**Competing interests**

The authors claim there is no competing interests.

**Materials & correspondence**

Correspondence and requests for materials should be addressed to Q.X. (XIANG.Qingyi@nims.go.jp) and S.M.(MITANI.Seiji@nims.go.jp).

# Supplemental materials

*1. PMA properties evaluation.*

A fully epitaxial MgO (5 nm)/Cr (30 nm)/Fe-QW (tFe: 3–7 ML)/MgAl$_2$O$_4$ (2 nm)/Fe (10 nm)/Ru (15 nm) prepared on a single crystalline MgO (001) substrate. Using the MR method mentioned in the main text, the PMA properties with different Fe thickness was evaluated. Figure S1 shows ***M-H*** curves with typical Fe thickness, with $t_{Fe}$ increasing, the PMA decreased a lot.

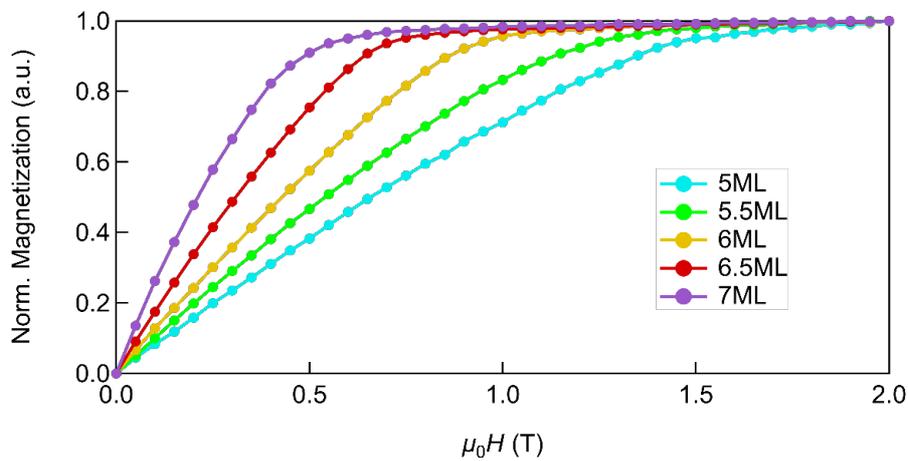

**Figure S1 M-H curves for different Fe thickness**

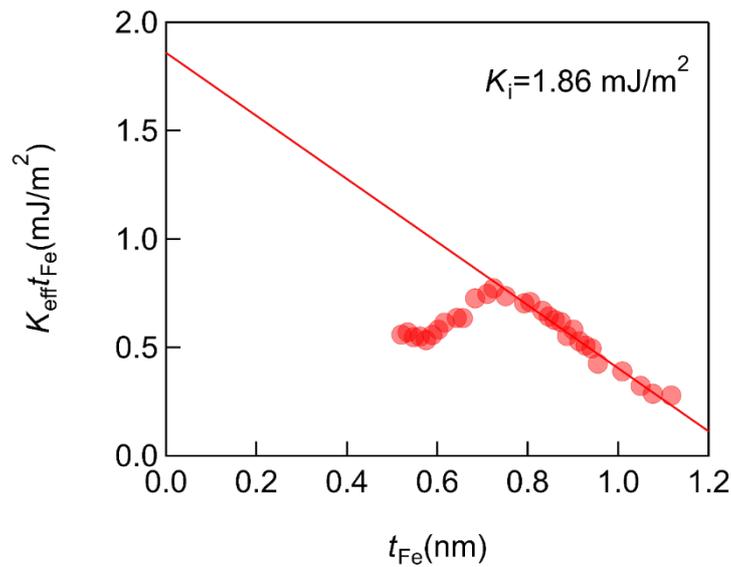

**Figure S2 $K_{eff}t_{Fe}$ as a function of $t_{Fe}$**

The PMA energies with different $t_{Fe}$ were summarized in Figure S2, where we assume the PMA energy following a simple equation as:[1]

$$K_{eff} = \frac{K_i}{t_{Fe}} - 2\pi M_s^2 + K_v \quad (1)$$

where $K_i$, $-2\pi M_s^2$, and $K_v$ are the interface, shape, and volume anisotropy energy densities, respectively. Here, $K_i$ is calculated around 1.86 mJ/m$^2$.

## 2. TEM image for thickness confirmation

A TEM observation is performed to obtain the interlayer structure of the prepared stacks. The TEM is taken at the place where the Fe thickness supposed to be 5-ML from the Fe-wedge sample. The supposed 5-ML position is determined by dI/dV spectrum. Figure S3 is the obtained TEM image, with the EDS mapping and RHEED pattern. Combining with these techniques, it is confidently to say perfect interfaces are prepared for Cr/Fe and Fe/MgAl$_2$O$_4$. And the EDS mapping also indicating that the position where determined as 5-ML by dI/dV measurement is indeed 5-ML of Fe atoms.

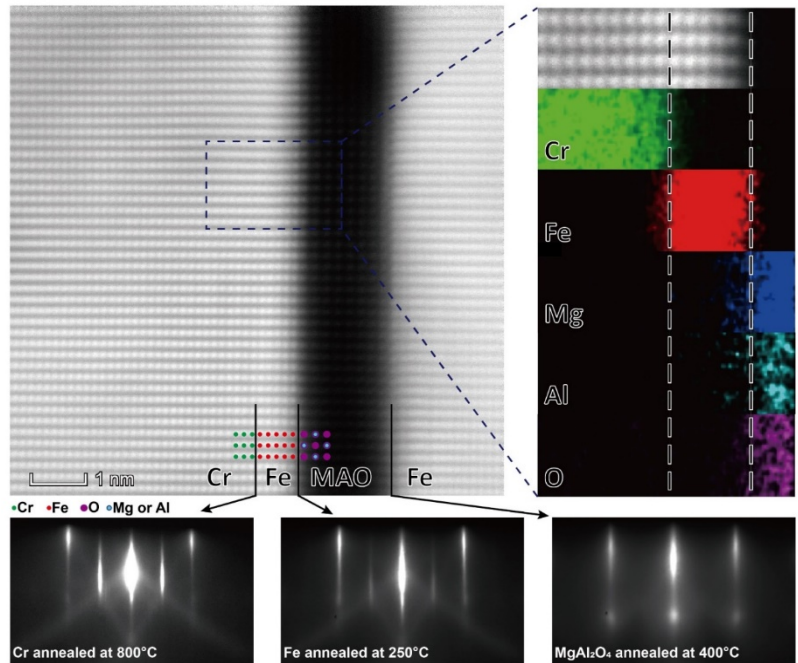

**Figure S3** TEM image taken for Fe/MgO/Fe multilayer structure where the position correlated to $t_{Fe}$= 5 ML.

## 3. Selected R-H curves under different bias voltage

The VCMA measurement are carried out using 5- and 6-ML samples. The MR curves are measured under different applied voltage for samples with 5- and 6-ML Fe as Figure S4 shows. It is clearly observed that saturation field is modified due to different applied voltage. Generally, negative applications of bias voltages cause increases of PMA while positive ones cause decreases.

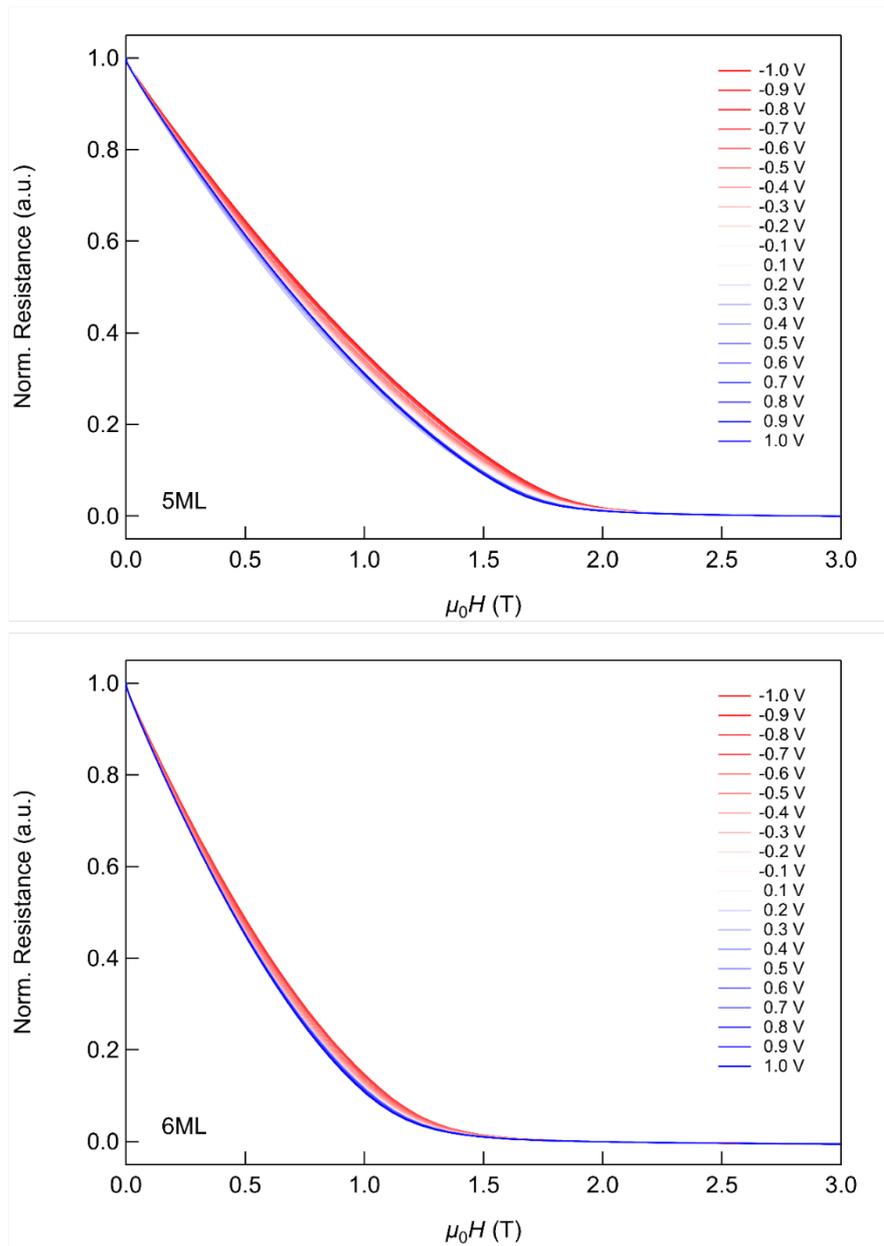

**Figure S4 typical R-H loops with different applied voltage for samples with 5- and 6-ML Fe**

Reference:

1. Johnson, M. T., Bloemen, P. J. H., Broeder, F. J. A. den & Vries, J. J. de. Magnetic anisotropy in metallic multilayers. *Rep. Prog. Phys.* **59**, 1409–1458 (1996).

2. Wang, Y., Lu, Z.-Y., Zhang, X.-G. & Han, X. F. First-Principles Theory of Quantum Well Resonance in Double Barrier Magnetic Tunnel Junctions. *Phys. Rev. Lett.* **97**, 087210 (2006).

3. Nozaki, T., Tezuka, N. & Inomata, K. Quantum Oscillation of the Tunneling Conductance in Fully Epitaxial Double Barrier Magnetic Tunnel Junctions. *Phys. Rev. Lett.* **96**, (2006).

4. Greullet, F. *et al.* Evidence of a Symmetry-Dependent Metallic Barrier in Fully Epitaxial MgO Based Magnetic Tunnel Junctions. *Phys. Rev. Lett.* **99**, 187202 (2007).

5. Iovan, A. *et al.* Spin Diode Based on Fe/MgO Double Tunnel Junction. *Nano Lett.* **8**, 805–809 (2008).

6. Nagahama, T., Yuasa, S., Suzuki, Y. & Tamura, E. Quantum-well effect in magnetic tunnel junctions with ultrathin single-crystal Fe(100) electrodes. *Appl. Phys. Lett.* **79**, 4381–4383 (2001).

7. Lu, Z.-Y., Zhang, X.-G. & Pantelides, S. T. Spin-Dependent Resonant Tunneling through Quantum-Well States in Magnetic Metallic Thin Films. *Phys. Rev. Lett.* **94**, 207210 (2005).

8. Niizeki, T., Tezuka, N. & Inomata, K. Enhanced Tunnel Magnetoresistance due to Spin Dependent Quantum Well Resonance in Specific Symmetry States of an Ultrathin Ferromagnetic Electrode. *Phys. Rev. Lett.* **100**, 047207 (2008).

9. Sheng, P. *et al.* Detailed analysis of spin-dependent quantum interference effects in magnetic tunnel junctions with Fe quantum wells. *Appl. Phys. Lett.* **102**, 032406 (2013).

10. Bauer, U., Przybylski, M. & Beach, G. S. D. Voltage control of magnetic anisotropy in Fe films with quantum well states. *Phys. Rev. B* **89**, 174402 (2014).

11. Chang, C.-H., Dou, K.-P., Guo, G.-Y. & Kaun, C.-C. Quantum-well-induced engineering of magnetocrystalline anisotropy in ferromagnetic films. *NPG Asia Mater.* **9**, e424 (2017).

12. Xiang, Q., Mandal, R., Sukegawa, H., Takahashi, Y. K. & Mitani, S. Large perpendicular magnetic anisotropy in epitaxial Fe/MgAl$_2$O$_4$(001) heterostructures. *Appl. Phys. Express* **11**, 063008 (2018).

13. Koo, J. W. *et al.* Large perpendicular magnetic anisotropy at Fe/MgO interface. *Appl. Phys. Lett.* **103**, 192401 (2013).

14. Yang, H. X. *et al.* First-principles investigation of the very large perpendicular magnetic anisotropy at Fe|MgO and Co|MgO interfaces. *Phys. Rev. B* **84**, 054401 (2011).